\documentclass[12pt,nohyper]{JHEP}
\usepackage{amsfonts}
\usepackage{amssymb}

\renewcommand{\Im}{{\rm Im}}
\renewcommand{\Re}{{\rm Re}}

\newcommand{\hf}{{\textstyle{1\over 2}}}
\newcommand{\be}{\begin{equation}}
\newcommand{\ee}{\end{equation}}

\newcommand{\ba}{\begin{eqnarray}}
\newcommand{\ea}{\end{eqnarray}}

\newcommand{\Xbar}{\overline{X}}

\newcommand{\xbar}{\overline{x}}

\newcommand{\tbar}{\overline{t}}
\newcommand{\taubar}{\overline{\tau}}
\newcommand{\lambar}{\overline{\lambda}}
\newcommand{\cF}{{\cal F}}
\newcommand{\cZ}{{\cal Q}}
\newcommand{\Fbar}{\overline{F}}
\newcommand{\Cbar}{\overline{C}}

\newcommand{\Omebar}{\overline{\Omega}}
\newcommand{\omebar}{\overline{\omega}}
\newcommand{\del}{\partial}
\newcommand{\delbar}{\overline{\del}}
\newcommand{\R}{{\mathbb{R}}}
\newcommand{\mZ}{{\mathbb{Z}}}

\newcommand{\Z}{{\cZ}}
\newcommand{\Zbar}{\overline{\cZ}}

\preprint{ITFA-2004-60\\hep-th/0412139\\}
\title{Attractors and The Holomorphic Anomaly}
\author{Erik Verlinde\\
Institute for Theoretical Physics, University of Amsterdam \\
Valckenierstraat 65, 1018 XE Amsterdam, The Netherlands}
\abstract{Motivated by the recently proposed connection between
$N=2$ BPS black holes and topological strings, I study the
attractor equations and their interplay with the holomorphic
anomaly equation. The topological string partition function is
interpreted as a wave-function obtained by quantizing the real
cohomology of the Calabi-Yau. In this interpretation the apparent
background dependence due to the holomorphic anomaly is caused by
the choice of complex polarization.  The black hole attractor
equations express the moduli in terms of the electric and magnetic
charges, and lead to a real polarization in which the background
dependence disappears. Our analysis results in a generalized formula
for the relation between the microscopic density of black hole
states and topological strings valid for all backgrounds.}

\begin{document}

\tableofcontents

\section{Introduction}
Recently an interesting connection has emerged between the counting of BPS black hole states 
and topological string theory. It was argued that the partition function of BPS black holes 
in Calabi-Yau compactifications of type II string theory is equal to the product of the 
partition sums of (anti-~) topological strings \cite{OSV}. As evidence it was shown that 
the corrected black hole entropy formula including its higher genus corrections
takes the form of a Legendre transform of the sum of the free
energies of the topological and anti-topological string. The
relevant higher genus correction had earlier been determined in
\cite{dewitetal}. A central role in the correspondence is played
by the black hole attractor equations \cite{FKS, strominger,
ferrara}, which express the values of the moduli fields at the
horizon in terms of the electric and magnetic charges of the black
hole. The proposed connection between BPS black holes and topological strings
has been further studied in special cases with few charges, see for example   
\cite{2dYM,atish,ulf,Sen}. In this paper I will discuss the generic situation with 
arbitrary number of charges in the context of type IIB string theory on a compact Calabi-Yau.

According to \cite{OSV} the mixed partition function associated
with an extremal black hole with magnetic charge $p^I$ and
electric potential $\phi^I$ factorizes as
\begin{equation}
\label{Zfact} Z_{bh}(p,\phi) =\left|\exp \cF_{top}\left(p+
{\textstyle{i\over 2}}\phi\right)\right|^2
\end{equation}
where $\cF_{top}$ denotes the free energy of the topological
string on the Calabi-Yau space $M$.
 It has the perturbative expansion
$$
{\cal F}_{top}(\lambda^{-1}X)=\sum_g \lambda^{2g-2}{\cal F}_g(X),
$$
where $\lambda$ is the string coupling constant and $X^I$ are the
periods of the holomorphic three form. In \cite{OSV} the following
identification was made,
\begin{equation}
\lambda^{-1}X^I=p^I+{\textstyle{i\over 2}}\phi^I.
\end{equation}
But there is a problem. The derivation of (\ref{Zfact}) as
presented in \cite{OSV} uses that ${\cal F}_{top}$ is holomorphic
in $X^I$. But it is well-known that the higher genus contributions
${\cal F}_g$ have an anholomorphic dependence on the complex
structure due to the holomorphic anomaly \cite{BCOV}. The problem
therefore is to reconcile the proposed relation between black hole
attractors and topological strings with the holomorphic anomaly.

Soon after the holomorphic anomaly was derived from the world
sheet formulation \cite{BCOV},  Witten interpreted the topological 
string partition as a wave function obtained by quantizing the space of 
three forms $H^3(M,\R)$ on $M$ in a complex polarization \cite{Witten}. In 
this interpretation the holomorphic anomaly describes the behavior  under
an infinitesimal change of the polarization, analogous to the way
the Knizhnik-Zamolodichikov equation 
appears in Chern-Simons theory. The wave function interpretation
of the holomorphic anomaly has been further
elaborated by Dijkgraaf, Vonk and the author \cite{DVV}.

In this paper I clarify the relation between the attractor
equations and the holomorphic anomaly by applying the results of
\cite{Witten, DVV}.  I'll arrive at a somewhat modified proposal
for the black hole partition function in terms of the topological
string, one that does not suffer from the holomorphic anomaly and
is background independent. The basic idea is as follows. Since the
holomorphic anomaly is the result of a background dependent choice
of complex polarization, one can remove it by using a background
independent real polarization. The attractor equations suggest a
real polarization, which instead of the complex moduli uses real
variables related to the electric/magnetic charges and potentials.

The outline of this paper is as follows.  I start with a review of
special geometry in section 2. In section 3 I introduce the
attractor equations and the entropy and free energy of BPS black
holes. The topological string partition function and the holomorphic anomaly
are described in section 4. In section 5 I discuss the
quantization of $H^3(M,\R)$ in a complex and in a real polarization. The topological string is shown define a background independent state.  The
application to the black hole partition function of BPS black 
holes described in section 6, where I'll present a formula for the number of 
BPS states.

\section{Special Geometry}

The $N=2$ supergravity theory corresponding to type $IIB$ string
theory compactified on a Calabi-Yau manifold $M$ contains complex
scalar fields $X^I$ and vector fields $A_\mu^I$ with
$I=0,\ldots,h_{2,1}=\dim H^{2,1}(M)$. The scalars $X^I$ are
identified with the periods of the holomorphic 3-form $\Omega$ on
$M$ along the non-intersecting $A$-cycles
\begin{equation}
\label{Xdef}\int_{A^I}\Omega=X^I.
\end{equation}
 The periods around the dual $B$-cycles
with $\#(A^I,B_J)=\delta^I{}_J$
\begin{equation}
\label{Fdef} \int_{B^J}\Omega = \partial_I{\cal F}_0(X)
\end{equation}
define a holomorphic function  ${\cal F}_0(X)$ homogeneous in $X$
with weight two. It represents the prepotential of the
supergravity theory, and also equals the tree level free energy of
the topological string. The first three derivatives of ${\cal
F}_0(X)$ are denoted by
\begin{equation}
\label{tauC} F_I\equiv\del_I{\cal F}_0 , \qquad
\quad\tau_{IJ}\equiv \del_I\del_J {\cal F}_0,\qquad\quad
C_{IJK}\equiv \del_I\del_J\del_K {\cal F}_0.
\end{equation}
The variables $X^I$ parametrize the complex structure on the
Calabi-Yau $M$.

The  moduli space $\cal M$ of complex structures is a special
K\"{a}hler manifold with K\"{a}hler potential
\begin{equation}
\label{Kahler} K= -\log i\left(\Xbar^I F_I -X^I\Fbar_I\right)
\end{equation}
Since overall scalings of $\Omega$ do not affect the complex
structure, the $X^I$ represent {\it projective} coordinates on
$\cal M$. By taking the 3-form $\Omega$ to vary holomorphically
with the complex structure, one can choose {\it local} coordinates
$t_i$ and $\tbar_i$ on $\cal M$ with $i=1,\ldots,h_{2,1}$, so that
$X^I$ and $F_I$ are holomorphic in $t_i$. In this paper I'll use
the local coordinates $t_i$ and $\tbar_i$ on $\cal M$ as well as
their projective counterparts $X^I$.

The freedom to scale $\Omega$ by a holomorphic function of the
moduli leads to the K\"{a}hler gauge transformations
\begin{equation}
\label{trafo} X^I(t)\to e^{f(t)}X^I(t), \qquad
\qquad\overline{X}^I(\tbar)\to
e^{\overline{f}(\tbar)}\Xbar^I(\tbar)
\end{equation}
and
\begin{equation}
\label{Ktrafo} K(t,\tbar)\to K(t,\tbar)-f(t)-\overline{f}(\tbar)
\end{equation}
Meaningful equations are either invariant or covariant with
respect to these transformations. Mathematically speaking $X^I$
and $F_I$ are holomorphic sections of a line bundle ${\cal L}$
over $\cal M$ defined by transition functions of the form (\ref{trafo}).
The three form $\Omega$ is also a section of $\cal L$. The line
bundle $\cal L$ comes equipped with a connection
$$
\nabla_i=\del_i+\del_iK,
$$
and its curvature is equal to the K\"{a}hler metric
$$
G_{i\overline{j}}=\del_i\delbar_{\overline{j}}K
$$
on $\cal M$. An important role is played by holomorphic three
point functions $C_{ijk}(t)$, which are obtained from the
$C_{IJK}$ in (\ref{tauC}) by contracting each index with
$\nabla_iX^I$. The K\"{a}hler potential, metric and three point
function can be expressed as\footnote{These relations can be
verified using the Riemann bilinear identity
$$ \int_M \alpha\wedge\beta=\sum_I\left\lbrack
\int_{A^I}\alpha\int_{B_I}\beta-\int_{B_I}\alpha\int_{A^I}\beta\right\rbrack
\qquad\qquad \alpha,\beta \in H^3(M)
$$}
\begin{eqnarray}
e^{-K}&=& i\int_M\Omebar\wedge\Omega, \nonumber\\
e^{-K}G_{i\overline{j}} & = & i\int_M\nabla_i\Omega\wedge
\overline{\nabla}_{\overline{j}}\Omebar\label{identities} \\
\nonumber C_{ijk} & = & i\int_M\nabla_i\Omega\wedge
D_j\nabla_k\Omega
\end{eqnarray}
Here the covariant derivative $D_i$ contains the usual christoffel
connection $\Gamma_{ij}^k$ as well as the term $\del_iK$. Given a
complex structure the third cohomology of the Calabi-Yau
decomposes as
$$
H^3(M)=H^{3,0}\oplus H^{2,1}\oplus H^{1,2}\oplus H^{0,3}.
$$
The covariant derivative $\nabla_i\Omega$ and their complex
conjugates form a basis of $H^{2,1}\oplus H^{1,2}$, which is
isomorphic to the cotangent space of $\cal M$. Further derivatives
of these three forms are therefore linearly related to the
$\Omega$ and its first derivatives.  The identities
(\ref{identities}) imply  that modulo exact terms one has
\begin{equation}
\delbar_{\overline{i}}\nabla_j\Omega=G_{\overline{i}j}\Omega,\qquad\quad
D_i\nabla_j\Omega=-e^K
C_{ij}{}^{\overline{k}}\,\overline{\nabla}_{\overline{k}}\Omebar
\label{Ovaria}
\end{equation}
These relations will be useful in explaining the origin of the
holomorphic anomaly. They further imply that the modified
connection that includes $e^KC_{ij}^{\overline{k}}$ in its
definition has zero curvature. This is a defining property of
special K\"{a}hler geometry.

\section{Black Hole Attractors}

In this section I describe the attractor equations, and review the
results of \cite{OSV} on the free energy of BPS black holes and
its connection with topological strings.

An extremal black hole with electric charges $q_I$ and magnetic
charge $p^J$ corresponds in the type $IIB$ theory to a three-brane
wrapping the three-cycle
\begin{equation}
\label{CC} {\cal C}=q_{I} A^I-p^J B_J
\end{equation}
on $M$. The BPS mass the black hole is expressed in terms of the
period integral of $\Omega$ around ${\cal C}$,
\begin{equation}
\label{Mbps} M^2_{BPS}=e^{K}\left|\Z\right|^2,
\end{equation}
with
\begin{equation}
\label{Qgravi} \Z = \int_{{\cal C}}\Omega=q_I X^I-p^I F_I.
\end{equation}
Here $K$ denotes the K\"{a}hler potential (\ref{Kahler}). The
linear combination (\ref{Qgravi}) of the electric and magnetic
charges $q_I$ and $p^J$ is known as the graviphoton charge. Notice
that the BPS mass (\ref{Mbps}) is invariant under the combined
K\"{a}hler transformations (\ref{trafo}) and (\ref{Ktrafo}), but
the graviphoton charge $\Z$ itself is not.

The complex scalar fields $X^I$ vary in general as a function of
the radial coordinate $r$ in the black hole geometry. For BPS
black holes these fields reach special values at the black hole
horizon characterized by the property that the expression
(\ref{Mbps}) is minimized. This phenomena is known as the
attractor mechanism. Equivalently the attractor values are
obtained by minimizing the graviphoton charge while keeping fixed
the K\"{a}hler potential. This leads to the condition
\begin{equation}
\label{one} q_I-\taubar_{IJ}p^J =i \lambda^{-1}
\left(F_I-\taubar_{IJ}X^J\right),
\end{equation}
where $\lambda^{-1}$ appears as a Lagrange multiplier. Its value
is determined by contracting both sides with $\Xbar^I$. This gives
\begin{equation}
\label{lambda} \lambda^{-1} =e^{K}\Zbar
\end{equation}
The complex equation (\ref{one}) is equivalent to two real
equations
\begin{equation}
{\Re}\Bigl(\lambda^{-1}X^I \Bigr)=  p^I, \qquad\qquad
{\Re}\Bigl(\lambda^{-1}F_I \Bigr)= q_{I}. \label{attrPQ}
\end{equation}
These are known as the attractor equations
\cite{strominger,ferrara}. The first equation is obtained by
taking the imaginary part of (\ref{one}). To find the second
equation one first multiplies (\ref{one}) by $\tau_{KI}$ before
taking again the imaginary part. The quantity $\lambda$ will in
the next sections be identified with the coupling constant of the
topological string. It behaves under K\"{a}hler transformation as
a section of $\cal L$, and hence the ratios $\lambda^{-1}X^I$ and
$\lambda^{-1}F_I$ that appear in the attractor equations are
invariant. Note further that the periods $X^I$ and $F_J$ as well
as the charges $p^I$ and $q_J$ depend on the choice of canonical
three cycles. Changing the choice of cycles transforms $(X^I,F_J)$
and $(p^I,q_J)$ by the same element of $Sp(2h_{2,1}\! +\! 2,\mZ)$.
The attractor equations (\ref{attrPQ}) are therefore
symplectically invariant.

The recent insight of \cite{OSV} is that the attractor equations
have a very natural thermodynamic interpretation.  The
Bekenstein-Hawking entropy $S_{bh}$ for a BPS black hole with
charges $q_I$ and $p^J$ is in leading order given by
\begin{equation}
\label{SXF} S_{bh} ={i\over 2} |\lambda|^{-2}\Bigl(
X^I\Fbar_I-\Xbar^I F_I\Bigr)
\end{equation}
where the graviphoton charge   $X^I$ and $F_I$ are evaluated at
the attractor point.
This expression can be written as
\begin{equation}
\label{Sformula} S_{bh}=\Im\Bigl( \lambda^{-2}X^IF_I\Bigr)
-2\Im\Bigl(\lambda^{-1}X^I\Bigr)\Re\Bigl(\lambda^{-1} F_I\Bigr)
\end{equation}
The first term is equal to $2\Im\,\cF_0(\lambda^{-1}X)$ due to the
homogeneity of $\cF_0$. Now remember that the real part of
$\lambda^{-1}X^I$ is set equal to the electric charge $p^I$ by the
first attractor equation.  I'll denote its imaginary part by $
\hf\phi^I$, so $X^I=p^I+{i\over 2}\phi^I$. The formula
(\ref{Sformula}) for $S_{bh}$ can then be recognized as the
Legendre transform of
\begin{equation}
F_{bh}(p,\phi)=2\Im\,\cF_0\left(p+{\textstyle {i\over 2}
}\phi\right).
\end{equation}
One has
\begin{equation}
S_{bh}(p,q)=F_{bh}(p,\phi) -\phi^I{\del
F_{bh}\over\del\phi^I}(p,\phi) \end{equation} where
\begin{equation}
q_I={\del F_{bh}\over\del\phi^I}(p,\phi)
\end{equation}
is just the second attractor equation. The interpretation of
$F_{bh}(p,\phi)$ is clear: it is the free energy associated with a
black hole with magnetic charge $p^I$ and electric potential
$\phi^I$.  The surprising fact is that $F_{bh}(p,\phi)$ is given
by the imaginary part of a holomorphic function of $p+{i\over 2}\phi$ equal
to the genus zero free energy of the topological string. This
observation could have been found many years ago, but only became
apparent when it was recognized this relation persists beyond
leading order.

Using the results of \cite{dewitetal} it was shown in \cite{OSV}
that to all orders the free energy is equal to the sum of the free
energies of the topological and anti-topological string. This was
taken as evidence for that fact that the black hole partition
function associated with a mixed ensemble with electric potentials
$\phi^I$ and fixed magnetic charges $p^I$
\begin{equation}
Z_{bh}(p,\phi) = \sum_{q}\Omega(p,q)e^{\phi\cdot q} \label{Zdef}
\end{equation} factorizes as
\begin{equation}
\label{Zbh}  Z_{bh}(p,\phi) =
\Bigl|\Psi_{top}(p+{\textstyle{i\over 2}}\phi)\Bigr|^2.
\end{equation}
Here $\Psi_{top}$ denotes the topological string partition
function. An interesting consequence is that the number of BPS
states $\Omega(p,q)$ may then be expressed as (ignoring some
factors of $2\pi$)
\begin{equation}
\label{Opsi}\Omega(p,q) = \int \!d\chi\, e^{-i\pi\, \chi q} \,\,
\Psi^*_{top} (p- \hf \chi) \Psi_{top}(p+ \hf \chi)
\end{equation}
This expression is just the Wigner function associated with
$\Psi_{top}(p)$ when regarded as a wave function. The aim of this
paper is to illuminate the wave function interpretation of the
topological string partition function, and also to see what one
can learn about topological string theory in general from this new
perspective.

A useful ingredient in connecting the partition functions of
topological strings and BPS black holes  is the following
description of the attractor equations in terms of $H^3(M,\R)$.
The 3-cycle ${\cal C}$ in (\ref{CC}) corresponding to a BPS black
hole with charges $p^I$ and $q_I$  is Poincare dual to a closed
three form $\gamma$ with real periods
\begin{equation}
\label{gperiod} \int_{A^I}\gamma =p^I\, ,\qquad\qquad
\int_{B_J}\gamma=q_J.
\end{equation}
Given a complex structure $\gamma$ can be decomposed in the basis
consisting of the holomorphic three form $\Omega\in H^{3,0}$, its
covariant derivatives $\nabla_i\Omega \in H^{2,1}$ and their
complex conjugates,
\begin{equation}
\label{gammaO} \gamma={1\over 2}\Bigl(
\lambda^{-1}\Omega+x^i\nabla_i\Omega+\xbar^i\overline{\nabla}_i\Omebar+\lambar^{\,-1}\Omebar\Bigr).
\end{equation}
By evaluating the periods of the r.h.s.\ and
comparing the result with (\ref{gperiod}) one finds the following
relation between $\lambda$ and $x^i$ and the periods
\begin{equation}
\label{modattr} \Re\Bigl(\lambda^{-1}X^I+x^i\nabla_iX^I\Bigr) =
p^I\, , \qquad\quad \Re\Bigl(\lambda^{-1}F_I+x^i\nabla_iF_I\Bigr)
=  q_I.
\end{equation}
The attractor equations are recovered by putting $x^i\!=\!0$, and
hence they determine the complex structure for which $\gamma$ has
components only in $H^{3,0}$ and $H^{0,3}$. This observation is
due to Moore \cite{moore}. In the following sections the
parameters $\lambda$ and $x^i$ get an interpretation in the
context of topological string theory: $\lambda$ becomes identified
with the perturbative coupling constant, while the $x^i$ turn in
to couplings to the physical operators.

\section{The Holomorphic Anomaly}

Topological string theory is defined perturbatively in terms of
its world sheet description as a topological sigma model obtained
by twisting an $N=2$ superconformal field theory coupled to
topological gravity \cite{wittentop}.  I 'll consider the
$B$-model, which is distinguished from the $A$-model by the way
the left and right moving $U(1)$ currents are used to twist the
theory.    In the $B$-model amplitudes depend on the complex
structure moduli of the Calabi-Yau space, but are independent of
the K\"{a}hler moduli.


The physical operators $O_i$ of the $B$-model are in one to one
correspondence with the $(2,1)$-forms on the target Calabi-Yau
space $M$, which are in turn related to the complex structure
deformations of $M$. Following standard string perturbation theory
genus $g$ amplitudes involving vertex operators $O_i$ are
expressed as integrated correlation functions: each operator is
integrated over the Riemann surface $\Sigma_g$ and subsequently
the entire correlation function is integrated over the moduli
space of genus $g$ surfaces \cite{BCOV}, see also
\cite{topstring}. By introducing couplings $x^i$ for the operators
$O_i$ one defines a generating function of all genus $g$
amplitudes
\begin{equation}
W_g(x;t,\tbar)=\Bigl\langle \exp\sum_i {x^i \int_{\Sigma_g}
O_i}\Bigr\rangle_{t,\tbar}
\end{equation}
The $n$-point amplitudes  are found by successive differentiation
w.r.t. $x^i$ and putting $x^i=0$ afterwards. The partition
function of the topological string has the perturbative expansion
\begin{equation}
\label{psitop} \Psi_{top}(x,\lambda ;t,\tbar)=\lambda^{{\chi\over
24}-1}\exp \sum_g \lambda^{2g-2}W_g(\lambda x; t,\tbar)
\end{equation}
where $\lambda$ is the coupling constant and $\chi$ is the Euler
number of $M$. For future convenience I inserted a factor
$\lambda$ in front of the $x^i$ so that  an $n$-point genus $g$
amplitude is weighed with $\lambda^{2g-2+n}$. Only positive powers
of $\lambda$ appear because amplitudes with $2g-2+n\leq 0$ vanish.
Due to the $\lambda$-dependent pre-factor $\Psi_{top}$ is a
section of ${\cal L}^{{\chi\over 24}-1}$.

To make the connection with the expression (\ref{Opsi})  the
topological string partition function $\Psi_{top}$ needs to be
converted from a function of the couplings $x^i$ and $\lambda$ to
a function of the charges $p^I$ and/or potentials $\phi^I$. An
important clue comes from the `modified attractor relations'
(\ref{modattr}). By themselves these are not yet sufficient,
because they involve too many variables: both $p^I$ and $q_I$
occur and also the couplings $x^i$ and $\lambda$ appear together
with their complex conjugates. As I'll explain in the following
section, the correct way to convert $\Psi_{top}$ to a function of
the charge $p^I$ is to regard $\Psi_{top}$ as a wave function and
to use (\ref{modattr}) to convert it from a complex polarization
to a real polarization. This same procedure will also remove the
background dependence on the complex structure moduli $t_i$ and
$\tbar_i$.

The moduli dependence of $\Psi_{top}$  is described by the
holomorphic anomaly equations. There are two such equations, one
describing the dependence on the anti-holomorphic moduli and the
other on the holomorphic moduli. They read
\begin{eqnarray}
\label{first} \overline{\del}_{\overline{i}}\Psi_{top}& = &
\left\lbrack{e^{2K}\over
2}\Cbar_{\overline{i}}{}^{jk}{\del^2\over\del x^{j}\del x^{k}}
+G_{ij}
x^j{\del\over\del\lambda}{}_{-1}\right\rbrack\Psi_{top}\\
\left\lbrack {\nabla_i}+\Gamma^k_{ij} x^j{\del\over\del x^k}
\right\rbrack\Psi_{top} &=& \left\lbrack
\lambda^{-1}{\del\over\del x^i}-{1\over 2}\del_i\log|G|-{1\over 2}
C_{ijk}x^jx^k\right\rbrack\Psi_{top}\label{second}
\end{eqnarray}
where the connection is given by
\begin{equation}
\mbox{$ $}\qquad \nabla_i=\del_i+\del_iK\left({h+1\over
2}+x^j{\del\over\del x^j} -\lambda{\del\over\del\lambda}\right)
\end{equation}
and $h=h_{2,1}$ and $|G|=\det G$. These equations were originally
derived by world sheet techniques, but, as we will see in the next
section, they have an interpretation that is quite independent of
that in terms of the quantization of $H^3(M.\R)$.

The holomorphic anomaly equations (\ref{first}) and (\ref{second})
differ slightly from those presented in \cite{BCOV}. Some small
differences are due to the factor $\lambda$ in front of $x^i$ in
(\ref{psitop}). A more substantial difference is that the second
anomaly equation given in \cite{BCOV} contains the genus one
amplitude $\cF_1$ instead of $\log|G|$ and also the coefficient in
front of $\del_iK$ in the connection is different. To arrive at
(\ref{second}) I eliminated $\cF_1$ using its the own anomaly
equation \cite{BCOV1}
\begin{equation}
\label{anholom} \delbar_{\overline{i}}\del_j{\cal
F}_1={e^{2K}\over 2}\Cbar_{\overline{i}}{}^{kl}
C_{klj}-\left({\chi\over 24}-1\right)G_{i\overline{j}}
\end{equation}
This equation is solved by
\begin{equation}
\label{F1} \cF_1={1\over 2}\log |G|+\left({h+1\over 2}-{\chi\over
24}+1 \right)K+f_1 +\overline{f}_1
\end{equation}
where  $f_1$ is holomorphic in the moduli. In fact,
(\ref{anholom}) is contained in (\ref{first}) by taking the linear
term in $x^i$.   The function $f_1$ is removed by multiplying
$\Psi_{top}$ with $\exp f_1$, which turns $\Psi_{top}$ in to a
section of ${\cal L}^{h+1\over 2}$ (times the square root of the
holomorphic determinant bundle) instead of ${\cal L}^{{\chi\over
24}-1}$. By combining these steps the second holomorphic anomaly
equation of \cite{BCOV} is turned in to (\ref{second}). An
important consequence of rewriting the second anomaly equation in
this way is that it is now defined without any reference to the
topological string partition function itself. Furthermore, both
holomorphic anomaly equations are now linear, and therefore the
set of solutions has become a linear space.

In the next section it will be shown that the linear space of
solutions of the holomorphic anomaly equations can be identified
with the space of wave functions obtained by quantizing
$H^3(M,\R)$ in a complex polarization. In particular, the
topological string partition function will be identified with a
state $|\Psi_{top}\rangle$. The second anomaly equation
(\ref{second}) is the conjugate of the first one (\ref{first})
with respect to the norm
\begin{equation}
\label{norm} 
\langle\Psi_{top}|\Psi_{top}\rangle =\int d\mu_{x,\lambda}
\exp\left\lbrack -e^{-K}G_{i\overline{j}}
x^i\xbar^{\overline{j}}+e^{-K}(\lambar\lambda)^{-1}\right\rbrack
\Bigl|\Psi_{top}(x,\lambda;t,\tbar)\Bigr|^2
\end{equation}
where
$$
d\mu_{x,\lambda}=  d^2\!\lambda\, d^{2h}x\, |\lambda|^{-4}
e^{{h+1\over 2} K} |G|^{1\over 4}
$$
It is straightforward to verify that by differentiating the r.h.s.
with respect to $t_i$ or $\tbar_i$ and using both holomorphic
anomaly equations that at least formally this norm is independent
of the background moduli. The significance of this fact will
become clear in the following.

\section{The Quantization of $H^3(M,R)$.}

The third cohomology $H^3(M,\R)$ has the structure of a phase
space: it describes the classical solutions for the 7 dimensional
action \cite{global, samson}, see also \cite{topm},
$$
S= \int_{M\times \R}\!\gamma \wedge d\gamma,
$$
where $\R$ corresponds to the `quantization time'. The equations
of motion imply that $\gamma$ is time independent and closed.  To
quantize $H^3(M,\R)$ one can use this action as a starting point,
or equivalently, its natural symplectic form
\begin{equation}
\label{Q} Q_{sympl}=\int_M d'\gamma\wedge' d'\gamma.
\end{equation}
Here $d'$ is an exterior derivative on the space of three forms,
and hence the wedge product $\wedge'$ on the r.h.s. of (\ref{Q})
is a combination of the one for three forms on $M$ as well as the
wedge product of $d'\gamma$ as a cotangent vector to  $H^3(M,\R)$.
At this point there are two natural ways to proceed: either one parametrizes $\gamma$ in 
terms of the couplings $\lambda$ and $x^i$ using the decomposition (\ref{gammaO}), or 
one uses its real periods $p^I$ and $q_J$ defined in (\ref{gperiod}). The first
choice naturally leads to quantization of $H^3(M,\R)$ in a complex
polarization, also known as K\"{a}hler quantization, while the
second choice gives a real polarization.

\subsection{K\"{a}hler quantization.}

To quantize $H^3(M,\R)$ in the complex polarization  one first chooses a 
complex structure on $M$. Next one decomposes $d'\gamma$ as in
(\ref{gammaO}) in terms of infinitesimal variations of the
couplings $\lambda^{-1}$ and $x^i$. Explicitely,
$$
d'\gamma={1\over 2}\Bigl(
d\lambda^{-1}\Omega+dx^i\nabla_i\Omega+d\xbar^{\overline{i}}\overline{\nabla}_{\overline{i}}\Omebar+d\lambar^{-1}\,\Omebar
\Bigr).
$$
Inserting this decomposition in (\ref{Q}) and by making
use of the relations (\ref{identities}) one finds for the
symplectic form
\begin{equation}
\label{Qcomplex} Q_{sympl}=ie^{-K}d\lambar^{-1}\!\!\wedge
d\lambda^{-1} +ie^{-K}G_{i\overline{j}}dx^i\!\wedge
d\xbar^{\overline{j}}.
\end{equation}
Generally, a symplectic form $Q_{ab}d\xi^a\wedge d\xi^b $ leads
after quantization to the commutation relations $ \bigl\lbrack
\xi^a,\xi^b\bigr \rbrack=iQ^{ab},$ where $Q^{ab}$ is the inverse
of $Q_{ab}$. In this case
\begin{equation}
\label{comm} \left\lbrack
\lambar^{-1},\lambda^{-1}\right\rbrack=e^K\qquad\qquad
\Bigl\lbrack x^i,\xbar^{\overline{j}}\Bigr\rbrack=e^K
G^{i\overline{j}}.
\end{equation}
These are the commutation relations of a set of creation and
annihilation operators.

One can now associate a state $|\Psi_{top}\rangle$ with the
topological string so that
\begin{equation} \label{topstate}
\Psi_{top}(x,\lambda;t,\tbar)=\langle\Psi_{top}|x,\lambda\rangle.
\end{equation}
where $|x,\lambda\rangle$ are the coherent eigenstates of $x^i$
and $\lambda$. These states depend on $t^i$ and $\tbar^i$, since
$\lambda^{-1}$ and $x^i$ are defined w.r.t. a reference complex
structure on $M$. In fact, all of the moduli dependence in the
wave function $\Psi_{top}$ is contained in these states; the state
$|\Psi_{top}\rangle$ itself is background independent. To derive
this result let me, following Witten \cite{Witten}, go back to the
decomposition (\ref{gammaO}). Since $\gamma$ itself is independent
of the moduli, the variations of $\lambda^{-1}$ and $x^i$ should
cancel those of the three forms $\Omega$ and $\nabla_i\Omega$. The
latter are given in (\ref{Ovaria}). In this way one finds
$$
\delbar_{\overline{i}}\lambda^{-1} =
-G_{\overline{i}j}x^j,\qquad\quad \delbar_{\overline{i}}x^k =
e^K\overline{C}_{\overline{i}\overline{j}}^k \xbar^{\overline{j}}
$$
A change of $\tbar_i$ therefore acts on the states as an
infinitesimal Bogolyubov transformation, since it mixes creation
and annihilation operators. After a little bit of puzzling one
finds that the coherent states indeed satisfy the holomorphic
anomaly equation
\begin{equation}
\label{anstate}
\overline{\del}_{\overline{i}}|x,\lambda\rangle=\left\lbrack{e^{2K}\over
2}\Cbar_{\overline{i}}{}^{jk} {\del^2\over\del x^{j}\del x^{k}}
-G_{\overline{i}j}
x^j{\del\over\del\lambda}{}_{-1}\right\rbrack|x,\lambda\rangle
\end{equation}
By commuting $x^i$ and $\lambda^{-1}$ through both sides of this
equation one easily verifies that with this variation the state
$|x,\lambda\rangle$ remains an eigenstate of these operators. This
confirms that $|\Psi_{top}\rangle$ is indeed independent of the
$\tbar$ moduli.

In a similar way one can show that it is also independent of
$t^i$. This fact also follows from the observation that the second
holomorphic anomaly equation (\ref{second}) is the conjugate of
the first (\ref{first}) with respect to the norm (\ref{norm}).
This implies that the inner product of $|\Psi_{top}\rangle$ with
any background independent state $|\Psi\rangle$ is also background
independent. This can only be true if $|\Psi_{top}\rangle$ itself
is independent of the background moduli.

The states $|x,\lambda\rangle$ are strictly speaking not part of
the Hilbert space. This is due to the fact that $\lambda^{-1}$
behaves like a creation operator instead of an annihilation
operator, and hence its coherent states are not normalizable. This
fact is also the cause of the upside-down upside down gaussian
integral over $\lambda^{-1}$ in the expression (\ref{norm}) for
the norm of $|\Psi\rangle_{top}$. To deal with this problem, let
me also introduce the operator $\cZ$ corresponding to the
graviphoton charge by writing $\lambar^{-1}\!=e^K{\cZ}$. Together
with its conjugate it satisfies
$$
\Bigl\lbrack \cZ,\overline{\cZ}\Bigr\rbrack=e^{-K}
$$
The Hilbert space ${\cal H}_{H^3(M,\R)}$ is spanned by the
normalizable coherent eigenstates $|x,\cZ\rangle$ of $x^i$ and
$\cZ$. These states satisfy
\begin{equation}
\label{inprod} \langle \xbar,{\overline{\cZ}}|y,{\cal Q}'\rangle=
\exp\left\lbrack
e^{-K}G_{\overline{i}j}\xbar^{i}y^j+e^K{\overline{\cZ}}{\cal
Q}'\right\rbrack \langle 0|0\rangle
\end{equation}
where $|0\rangle$ denotes the ground state. The coherent states
$|x,\lambda\rangle$ can now be represented as formal integrals
\begin{equation}
\label{coherent} |x,\lambda\rangle =\int\!
d{\cZ}\,\exp\left(-\lambda^{-1}{\cZ}\right)\,|x,{\cZ}\rangle
\end{equation}
The norm (\ref{norm}) of $|\Psi_{top}\rangle$ can thus be
reexpressed in terms of the overlaps with normalizable states
$|x,\cZ\rangle$. These obey the completeness relation
\begin{equation}
\label{complete} 1\!\! 1=\int\!d\mu_{x,{\cZ}}\, \exp\left\lbrack
-e^{-K}G_{ij}\xbar^{i}x^j-e^K{\overline{\cZ}}{\cZ}\right\rbrack
|x,{\cZ}\rangle\langle \xbar,{\overline{\cZ}}|
\end{equation}
where\footnote{Consistency of (\ref{inprod}) with (\ref{norm})
leads to $\langle 0|0\rangle =\exp{h+1\over 2}K |G|^{1\over 4}$}
$$
d\mu_{x,{\cZ}}=  d^2\!{\cZ}\, d^{2h}x\,  {e^{{h-3\over 2}
K}|G|^{1\over 4}}
$$
The perturbative expansion of the overlaps
$\langle\Psi_{top}|x,{\cZ}\rangle$ is given by the Borel transform
of the original expansion, and hence is possibly convergent. This
is an indication that the state $|\Psi_{top}\rangle$ is not only
background independent, but also normalizable.

\subsection{Real polarization}

The third cohomology $H^3(M,\R)$ is defined in pure topological
terms, and can be quantized in a background independent fashion by
using the real periods $p^I$ and $q_J$ of $\gamma$ instead of the
complex couplings. Using the Riemann bilinear identity one 
writes the symplectic form $Q_{sympl}$ in terms of the periods of the variation $d'\gamma$,
$$
\int_{A^I}d'\gamma =dp^I\,
,\qquad\qquad \int_{B_J}d'\gamma=dq_J,
$$
 as
\begin{equation}
\label{Qreal} Q_{sympl}=\sum_I dp^I\wedge dq_I.
\end{equation}
Here one recognizes the standard symplectic form on the phase
space of $q$'s and $p$'s. The real periods $q_I$ and $p^J$ turn
after quantization in to hermitean operators with the standard
commutation relations
\begin{equation}
\label{pq} \Bigl\lbrack p^I,q_J\Bigr\rbrack= -i \delta^I{}_J
\end{equation}
The periods $p^I$ and $q_J$ are related to the complex couplings
$\lambda$ and $x^i$ through the equations (\ref{modattr}). It is
possible to verify that these indeed represent a canonical change
of variables. The Hilbert space is spanned by the eigenstates of
$q_I$ or $p^I$. For definiteness, I'll work with the eigenstates
$|p\rangle$ of the latter.

The wave functions obtained in the K\"{a}hler quantization can be
converted to more standard wave functions in the real variables
$p^I$ or $q_I$. In particular, the state $|\Psi_{top}\rangle$
associated with the topological string can now be represented as a
background independent wave function
\begin{equation}
\Psi_{top}(p)=\langle \Psi_{top}|p\rangle
\end{equation}
Locally, the states $|p\rangle$ are indeed independent of the
complex structure moduli. However, because $p^I$ and $q_I$ are
defined w.r.t. a choice of 3-cycles, their eigenstates have
non-trivial monodromy properties. Suppose a 3-cycle, say
$C_{m,n}=m_IA^I-n^IB_I$, shrinks to zero size at a special locus
in $\cal M$. By taking a closed path around its zero locus the
other cycles pick up a monodromy
\begin{equation}
A^I\to A^I+n^IC_{m,n}, \qquad\qquad B_I\to
B_I+m_IC_{m,n}.\label{monod}
\end{equation}
The periods $p^I$ and $q_I$ transform accordingly, that is
\begin{equation}
\label{monodromy} p^I \to p'^I=
p^I\!+\!n^I(m_Jp^J\!-\!n^Jq_J),\qquad q_I \to q'_I= q_I\! + \!
m_I(m_Jp^J \!-\! n^Jq_J).
\end{equation}
This makes clear that the effect of the monodromy is that the
states $|p\rangle$ are changed by the unitary transformation
\begin{equation}
\label{generators} |p\rangle\to \exp {i\over
2}(m_Ip^I-n^Iq_I)^2|p\rangle
\end{equation}
Similar monodromies are picked up at all the loci at which a
3-cycle shrinks. Together they generate the modular group of the
Calabi-Yau space $M$, which is a subgroup of $Sp(2h+2,\mZ)$. I
will assume that the state $|\Psi_{top}\rangle$ is monodromy
invariant. The wave function $\Psi_{top}(p)$, however, is in
general not modular invariant but transforms under the modular
group by canonical transformations.
The significance of this observation will become clear in the
next section.

\subsection{The big phase space}

To express  the original topological partition function in terms
of  $\Psi_{top}(p)$, and vice versa, it is convenient to work in
the `big phase space' \cite{DVV}. Instead of parametrizing the
complex structure in terms of the local coordinates $t_i$ and
$\tbar_i$, one uses the projective coordinates $X^I$ and its
conjugate. Furthermore, the couplings $x^i$ and $\lambda$ are
combined in to the variables
\begin{equation}
\label{xbig} x^I=\lambda^{-1}X^I+x^i\nabla_iX^I.
\end{equation}
These represent the coefficients of the decomposition
$\gamma={1\over 2}(x^I\omega_I+\xbar^I\omebar_I)$ in terms of the
basis $\omega_I =\del_I\Omega $ of $H^{3,0} \oplus H^{2,1}$. The
relation between $x^I$ and the real variables $p^I$ and $q_I$ is
again determined by evaluating the periods of $\gamma$. This
gives\footnote{The period integrals of the forms $\omega_I$ are
$$
\int_{A^I}\omega_J=\delta^I{}_J\qquad\qquad\int_{B_I}\omega_J=\tau_{IJ}.
$$}
\begin{equation}
\Re\left(x^I\right)=p^I\qquad\quad\Re\left(\tau_{IJ}x^J
\right)=q_I \label{miniattr}
\end{equation}
These equations are equivalent to (\ref{modattr}), and are solved
explicitly by
\begin{equation}
\label{two}
x^I=-i\left(\Im\tau^{-1}\right)^{IJ}\left(q_J-\taubar_{JK}p^K\right).
\end{equation}
The equations (\ref{miniattr}) and (\ref{two}) should be regarded
as operator identities. Therefore, one can take their expectation
values or compute their matrix elements. For example, for the
ground states $|0,{\cZ}\rangle$ this leads to the interesting
property  that the normalized expectation values of the charge
operators $p^I$ and $q_I$ obey the attractor equations
\begin{equation}
\Bigl \langle p^I\Bigr\rangle
=\Re\left(\lambda^{-1}X^I\right),\qquad\qquad \Bigl\langle
q^I\Bigr\rangle=\Re\left(\lambda^{-1}F_I\right),
\end{equation}
where $\lambda^{-1}=e^K\overline{\cZ}$. This follows from the fact
that the states $|0,{\cZ}\rangle$ are annihilated by $x^i$ and
hence the expectation value of $x^I$ is equal to
$\lambda^{-1}X^I$.

From the canonical commutators (\ref{pq}) and (\ref{two}) it is
clear that $x^I$ and its conjugate satisfy
$$
\Bigl\lbrack x^I,\xbar^J\Bigr\rbrack=
2\left(\Im\tau^{-1}\right)^{IJ}.
$$
This commutator is equivalent to those given in (\ref{comm}) for
the couplings $x^i$ and $\lambda^{-1}$. The states
$|x,\lambda\rangle$ can thus be identified with the coherent
eigenstates $|x\rangle$ of $x^I$. The completeness relation for
these states is
\begin{equation}
\label{compl} 1\!\!1=\sqrt{|\Im\tau|}\int \! dx\,d\xbar\,\,
\,e^{-{1\over 2} x^I\xbar_I}\, |x\rangle\langle
\xbar|,
\end{equation}
where $|\Im\tau|$ denotes the determinant of $\Im\tau_{IJ}$. Here
and in the following indices are raised and lowered using
$\Im\tau_{IJ}$ as the metric on the big phase space.

To be able to convert wave functions from the complex to the real
polarization one needs the overlaps $\langle p|x\rangle$. By
requiring that the relations (\ref{miniattr}) hold as operator
identities when inserted in between the two states $\langle p|$
and $|x\rangle$ one finds \cite{DVV}
\begin{equation}
\label{px} \langle p|x\rangle= \exp{i\over
2}\left(p^I\taubar_{IJ}p^J-p^Ix_I+\hf x^Ix_I\right).
\end{equation}
It thus follows that the original partition function can be
expressed in terms of $\Psi_{top}(p)$ by
\begin{equation}
\label{Psix} \Psi_{top}(x)=\int\!  dp\, \Psi_{top}(p)\,
\exp{i\over 2}\bigl(p^I\taubar_{IJ}p^J-p^Ix_I+\hf x^Ix_I\bigr).
\end{equation}
The inverse relation expressing $\Psi_{top}(p)$ in terms of
$\Psi_{top}(x)$ is easily obtained with the help of the
completeness relation (\ref{compl}). Finally, from the expression
(\ref{Psix}) it is almost manifest that the partition function
safisfies the holomorphic anomaly equations. In the big phase
space these takes the form
$$
 \overline{\del}_I \Psi_{top} = {1\over
2}\Cbar_{I}{}^{JK} {\del^2\over\del x^J\del x^K}\Psi_{top},
$$
$$
\left\lbrack \del_I+C^K_{IJ}x^J{\del\over\del x^K}
\right\rbrack\Psi_{top}  =  \Bigl\lbrack-{1\over
2}\del_I\log|\Im\tau|-{1\over 2}C_{IJK}x^Jx^K\Bigr\rbrack
\Psi_{top}
$$
For more details on the `big phase space version' of the
holomorphic anomaly equation, see \cite{DVV}.

\section{BPS States and Topological Strings}

Much of the material presented in the previous sections is in some
form or another contained in or scattered throughout the
literature. I have gone through it in detail to make the paper
self-contained, fill in gaps and to highlight the interplay
between the attractor equations and the holomorphic anomaly. But
now that the stage is set, let me return to discussing the
proposed relation between BPS black holes and topological strings.

Topological string theory has at present only a perturbative
formulation. In this respect the correspondence with the BPS black
holes is particularly interesting because it may provide
 non-perturbative formulation. It also strongly suggests
that such a formulation involves both  topological and
anti-topological strings. In this section I take the perspective
that the counting of BPS black hole states serves as a {\it
definition} of non-perturbative topological string theory.

Let $\Omega(p,q)$ be the number of black hole states counted with
weight $(-1)^F$. One can define an operator in the Hilbert space
obtained by quantizing $H^3(M,\R)$ by
\begin{equation}
\label{hatO} \hat{\Omega}=\sum_{p,q}\int \!d\chi\, e^{i\chi q}\,
|p- \hf\chi\rangle \, \Omega(p,q)\,\langle p+\hf\chi|
\end{equation}
where $|p\pm\hf\chi\rangle$ are the states obtained in the real
polarization. The relation  (\ref{hatO}) can be inverted to
\begin{equation}
\label{Oinv}\Omega(p,q) =  \int\! d\chi\, e^{-i\chi q} \,\,
\langle p- \hf \chi|\hat{\Omega}|p+ \hf \chi\rangle.
\end{equation}
This equation holds independent of the connection with topological
strings. The expression (\ref{Opsi}) described in section 3 gives
the relation between the number of BPS states $\Omega(p,q)$ and
the topological string in terms of the real polarization. The
formula (\ref{Opsi}) implies that the operator $\hat{\Omega}$
factorizes as
\begin{equation}
\label{Op} \hat{\Omega}=|\Psi_{top}\rangle\langle\Psi_{top}|
\end{equation}
In other words, $\hat{\Omega}$ is (proportional to) a projection
operator on the state $|\Psi_{top}\rangle$. In fact, below I will
argue that this relation is presumably only approximately true.
Namely, the r.h.s. of (\ref{Op}) should probably be generalized to
include a sum over states, so that $\hat{\Omega}$ behaves more
like a density matrix. Independent evidence for this fact is given
in \cite{2dYM} and \cite{Vafaetal}.

The results of the previous section can now be used to convert the
expression (\ref{Oinv}) from the real to the complex polarization
by writing it in terms of the coherent states $|x\rangle$ using
the overlap (\ref{px}). The key identity is
\begin{equation}
\label{key} \int \!d\chi\,e^{-iq\chi}\,\langle
x|p+\hf\chi\rangle\langle p-\hf \chi|\xbar\rangle={e^{-{1\over
2}|\!|x-x_{p,q}|\!|^2}\over\sqrt{|\Im\tau|}}
\end{equation}
where $x_{p,q}$ are the solutions to (\ref{miniattr}) given in
(\ref{two}), and the norm in the exponential is defined by
\begin{equation}
|\!|x-x_{p,q}|\!|^2\equiv
\Im\tau_{IJ}\left(x^I-x^I_{p,q}\right)\left(\xbar^J-\xbar_{p,q}^J\right)
\end{equation}
This norm  is not positive definite, however, since $\Im\tau_{IJ}$
has one negative eigenvalue corresponding to $x^I\sim X^I$. This
problem can be solved along the lines discussed in section 4.1,
but for simplicity I'll ignore this issue in the remainder.

Assuming that  (\ref{Op}) holds the matrix element $\langle
\xbar|\hat{O}|x\rangle$ factorizes as a product of $\Psi_{top}(x)$
and its conjugate.  Using (\ref{hatO}) and (\ref{key}) one thus
finds
\begin{equation}
\label{psps} \Psi_{top}^*(\xbar)
\Psi_{top}(x)=\sum_{p,q}\Omega(p,q){e^{-{1\over 2}
|\!|x-x_{p,q}|\!|^2}\over\sqrt{|\Im\tau|}}
\end{equation}
The r.h.s. defines a new partition function of BPS black holes
corresponding to a particular ensemble with a statistical weight
given by gaussian factor. This ensemble is apparently the natural
one from the point of view of the topological string. It would be
interesting to know what is special about this ensemble.

The expression for the $\Omega(p,q)$ in terms of the topological
partition function in complex polarization is obtained by
inverting the relation (\ref{psps}). Alternatively, one can start
from (\ref{Opsi}) and use the completeness relations
(\ref{compl}). Both methods yield the following result
\begin{equation}
\label{result} \Omega(p,q) =  \sqrt{|\Im\tau|}\int \! dx d\xbar\,
\,e^{-{1\over 2} |\!|x|\!|^2} \Psi_{top}^*(\xbar_{p,q}-\xbar)
\Psi_{top}(x_{p,q}+x)
\end{equation}
The reversal of the sign in front of $x$ is necessary so that the
gaussian integration is (mostly) convergent. Again to get a
completely convergent result one has to change to the normalizable
polarization in terms of $\cZ$.

The formula (\ref{result}) is one of the main results of this paper. It is
written in a short hand notation to make it readable. But hidden
in the wave functions $\Psi(x)$ is still the dependence on the
background moduli. The full expression is nevertheless independent of the choice of
background moduli. Hence, one is free to choose the background at
special points. For example, by sending the moduli to `infinity' in
such a way that $\Im\tau_{IJ}\to\infty$ one recovers the expression (\ref{Opsi}) 
as a special case .
Alternatively, one can for instance choose the moduli to be at the
attractor point. In that case the value of $x_{p,q}$ is equal to
the solutions $X_{p,q}$ of the attractor equations. By expanding
$\Psi_{top}$ and its conjugate around the background one can
derive a perturbative expressions for $\Omega(p,q)$ in terms of
the correlation functions of the topological string. I'll leave it
for future work to analyze the resulting expressions in detail.

Let me end with a few comments and open problems. First of all,
from the result (\ref{result}) it is not immediately clear that
$\Omega(p,q)$ are integers. Indeed, even the integral nature of
the charges $p^I$ and $q_J$ has hardly been used in this paper.
This brings us to a final comment regarding the factorized form
(\ref{Op}) of $\hat{\Omega}$. Without claiming to have a complete
understanding of this point, I believe that it is unlikely that
this identity holds exactly. This has to do with the following 
important property that is expected to hold for the number of BPS
states $\Omega(p,q)$. 

Since $\Omega(p,q)$ is integer it should
indeed be independent of small changes of the background complex
structure. Now, one can perform a monodromy that changes the
choice of basis of three cycles, and hence also the definition of
the charges by the monodromy transformations (\ref{monodromy}).
From this one concludes that $\Omega(p,q)$ should be monodromy
invariant: that is
\begin{equation}
\Omega(p,q)=\Omega(p',q'),
\end{equation}
where $p'$ and $q'$ are given in (\ref{monodromy}). This is
clearly a non-trivial property, comparable to the restriction of
modular invariance for 2d conformal field theory.  It implies that
the operator $\hat{\Omega}$ commutes with the generators
(\ref{generators}) of the monodromy transformation on the states.
Now, if $\hat{\Omega}$ indeed factorizes as in (\ref{Op}) this
would mean that the state $|\Psi_{top}\rangle$ itself is
invariant. In other words
\begin{equation}
e^{ {i\over 2}(m\cdot p-n\cdot
q)^2}|\Psi_{top}\rangle=|\Psi_{top}\rangle
\end{equation}
for all cycles $m_IA^I-n^IB_I$ that can shrink to zero. This is
analogous to demanding modular invariance of a conformal block,
which is generally only possible for very special CFT's. Clearly
this point needs further investigation.

Another more intuitive reason why the factorized form (\ref{Op})
is not likely to hold exactly is that it would mean that the
topological string by itself has a non-perturbative definition.
But the correspondence with the BPS black holes seems to suggest
that the non-perturbative definition involves both topological as
well as anti-topological strings. The formula (\ref{result})
should then be modified so that there is a sum over states on the
r.h.s. So one gets a density matrix formulation of the
non-perturbative theory defined by the numbers $\Omega(p,q)$. This
point of view will be further elaborated in forthcoming work
\cite{inprogress}.

\vspace{2cm}

\begin{flushleft}
{\bf Acknowledgements}
\end{flushleft}

I like to thank Annamaria Sinkovics, Robbert Dijkgraaf, Jan de
Boer, Greg Moore, Cumrun Vafa, Hirosi Ooguri, Herman Verlinde, Al
Shapere, John McGreevy, Mina Aganagic, Sergei Gukov and Andy
Neitzke for interesting and helpful dicussions. I also like to
thank the Aspen Center for Theoretical Physics, where part of this
work was done, for its hospitality and for providing a stimulating
atmosphere.

\end{document}